# Towards Integrated Alignment

**Authors:** Ben Y. Reis[1,2,3,4,5,6] and William La Cava[1,2]

1. Computational Health Informatics Program, Boston Children's Hospital, Boston, Massachusetts
2. Department of Pediatrics, Harvard Medical School, Boston, Massachusetts
3. Ivan and Francesca Berkowitz Living Laboratory, Harvard Medical School and Clalit Research Institute
4. Affiliated Faculty, Department of Biomedical Informatics, Harvard Medical School, Boston, Massachusetts
5. Harvard Data Science Initiative, Cambridge, Massachusetts
6. Faculty Associate, Berkman Klein Center for Internet and Society at Harvard University

**Abstract**

As AI adoption expands across human society, the problem of aligning AI models to match human preferences remains a grand challenge. Currently, the AI alignment field is deeply divided between behavioral and representational approaches, resulting in narrowly aligned models that are more vulnerable to increasingly deceptive misalignment threats. In the face of this fragmentation, we propose an integrated vision for the future of the field. Drawing on related lessons from immunology and cybersecurity, we lay out a set of design principles for the development of *Integrated Alignment* frameworks that combine the complementary strengths of diverse alignment approaches through deep integration and adaptive coevolution. We highlight the importance of *strategic diversity* - deploying orthogonal alignment and misalignment detection approaches to avoid homogeneous pipelines that may be "doomed to success". We also recommend steps for greater unification of the AI alignment research field itself, through cross-collaboration, open model weights and shared community resources.

# 1. Introduction

Aligning models to conform with human preferences and expectations is a central challenge for the future of AI [1]. Misalignments can emerge along several dimensions, including truthfulness, safety, ethics, and logical soundness, among others [2–5]. Detecting these misalignments becomes increasingly difficult as models scale in size and complexity, with some deceptive forms of misalignment undermining attempts to detect them [6–8]. There is an urgent need to develop a deeper understanding of emerging misalignment threats, alongside improved methods for identifying and correcting them.

The nascent field of AI alignment [9,10] explores a diverse range of approaches for aligning AI models and detecting misalignments, each with its own strengths and weaknesses. These approaches can be broadly divided into *behavioral* approaches that examine a model's inputs and outputs, and *representational* approaches that examine a model's inner workings [11,12]. Thus far, the vast majority of efforts have focused on only one of these approaches, leaving AI models potentially more vulnerable to a wide range of misalignment threats. The AI alignment field itself is also deeply split along this *behavioral-representational divide*, with limited communication between the two research communities [10,13–17].

In response to these growing challenges, we call for the development of *Integrated Alignment (IA)* frameworks that combine the complementary strengths of diverse alignment approaches with the aim of more robustly identifying a wide range of misalignments. Drawing practical lessons from the related fields of immunology and cybersecurity, we propose a set of design principles for IA frameworks. We highlight the importance of *strategic diversity* - deploying orthogonal forward and backward alignment approaches to mitigate the pitfalls of homogeneous alignment pipelines that may be "doomed to success." We also call for greater unification of the AI alignment field itself, encouraging communication across subspecialties, greater availability of open model weights, and a growing ecosystem of shared community resources.

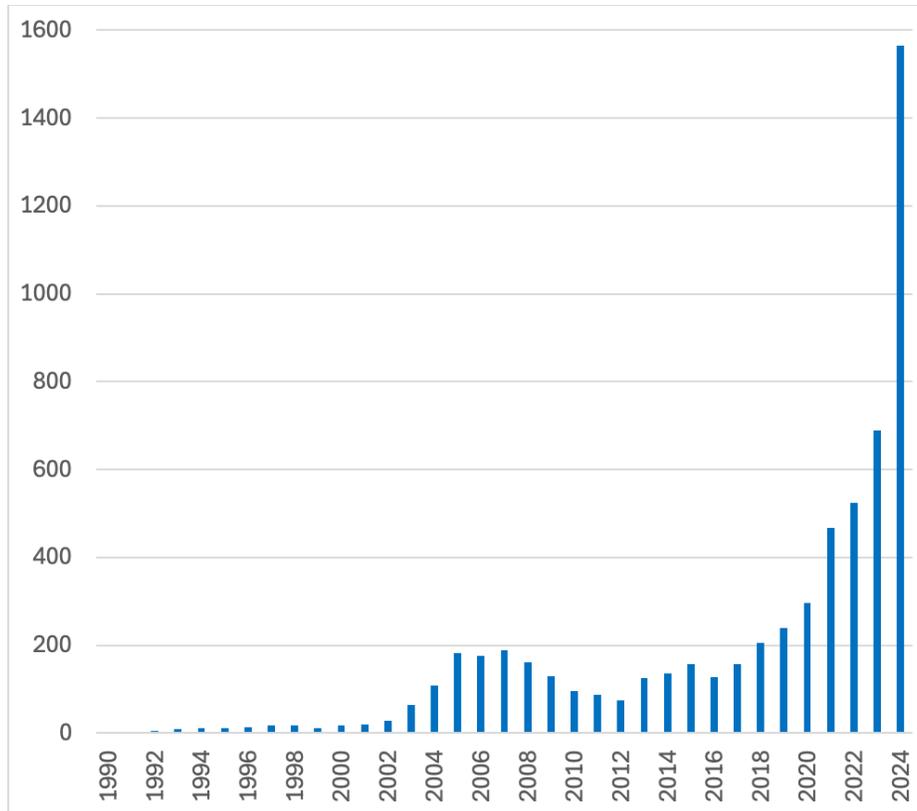

**Figure 1.** Annual number of publications returned for the search string "Artificial Intelligence alignment", 1990-2024 (Source: PubMed database).

## 2. AI Alignment: The Behavioral-Representational Divide

The field of AI alignment has grown exponentially in recent years (Figure 1). While a comprehensive overview of developments in this wide-ranging field [10,16] is outside the scope of this perspective, we provide a brief summary of some of the major approaches to alignment and misalignment detection. Several categorization schemes for the field have been proposed, including *backward alignment* vs. *forward alignment*[10], *outer alignment* vs. *inner alignment* [18,19], and others. In this Perspective, we focus on a central distinction that divides the field today: *behavioral alignment* vs. *representational alignment* (Figure 2).

Most alignment efforts to date have focused exclusively on analyzing either model behavior or internal model representations. Such narrow alignment approaches can leave models more vulnerable to a wide range of emerging misalignment threats.[16,20] Zhang et al. point out that approaches to understanding model behavior based either on model inputs or model internals are "studied and applied rather independently, resulting in a fragmented landscape of approaches and terminology" [15] Bereska and Gavves note that mechanistic interpretability has developed as a separate research area from behavioral approaches, with "diverging terminology" that "inhibits collaboration across disciplines." [16] This division is exacerbated by the fact that many developers of frontier AI models do not provide open access to their model weights, leaving behavioral approaches as the only usable option for alignment researchers. [20] Burden et al. note that even within the behavioral alignment field, "divergent evaluation paradigms have emerged, often developing in isolation, adopting conflicting terminologies, and overlooking each other's contributions. This fragmentation has led to insular research trajectories and communication barriers." [17]

We begin by summarizing the approaches on either side of this divide, including how they detect and correct misalignment, and their relative strengths and weaknesses. While the examples below relate to text-based LLMs, the challenges apply equally to models dealing with image, video, and other data modalities, as well as to agentic and multi-agent AI systems.

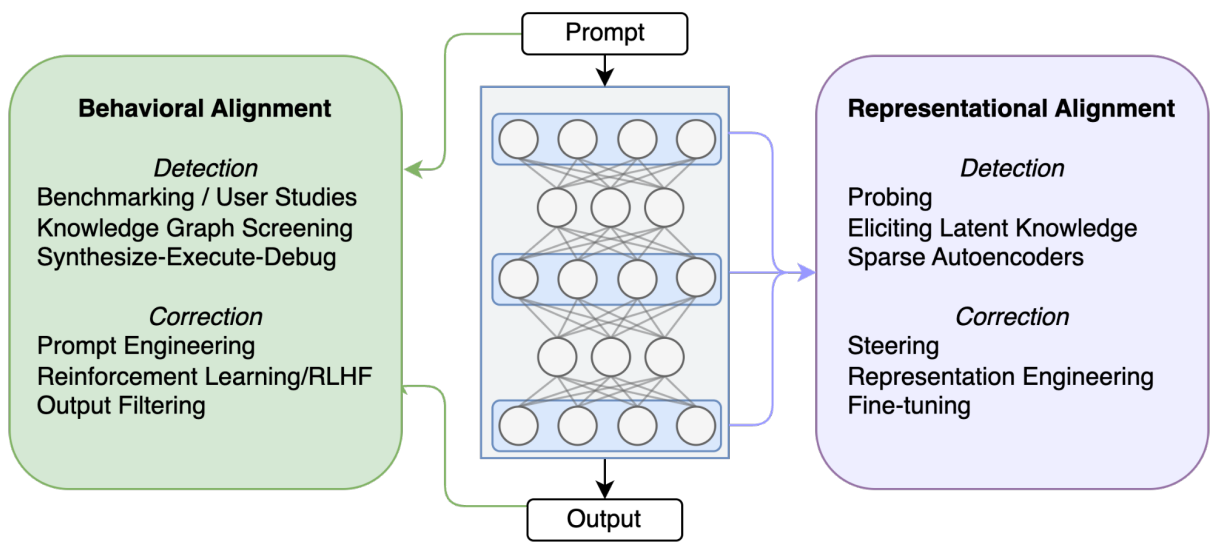

**Figure 2.** Behavioral alignment approaches focus on a model's inputs and outputs, whereas representational approaches focus on a model's internal activations and representations. (A small sample of detection and correction methods are shown here; for a more complete listing, see [10,16].) The proposed Integrated Alignment framework combines these approaches by examining model inputs and outputs alongside internal activations and representations.

## *Behavioral Alignment*

Behavioral approaches seek to align a model in a "black-box" fashion – based only on its inputs and outputs, without access to the activations or representations in the intervening layers.

### *Detection of Misalignment*

Behavioral approaches to misalignment detection can take the form of benchmarks or standardized exams[21], which can sometimes be limited in their translation to real-world tasks[22]. They can also be implemented via user studies involving domain experts or end-users of an application [23,24]. Researchers have proposed additional behavioral alignment detection methods for specific domains that involve more sophisticated processing of outputs. For example, Alber et al. [25] proposed extracting biomedical concepts and relations from model responses and verifying them against a biomedical knowledge graph. In the field of software code generation, the "synthesize-execute-debug" approach takes model outputs in the form of code and compiles, executes and debugs them to identify misalignments [26,27]. Mathematical and other reasoning tasks may also employ automated verification algorithms such as theorem checkers[28].

*Correction of Misalignment*

Many behavioral approaches to misalignment correction have been explored, the most popular being Reinforcement Learning with Human Feedback (RLHF) [29]. In this approach, used to train the original InstructGPT model[29], human labelers indicate their preferences among several model outputs and provide sample output demonstrations which are then used for fine-tuning. Additional Behavioral Alignment approaches include Iterated Distillation and Amplification (IDA)[30], Recursive Reward Modeling (RRM)[31], Cooperative Inverse Reinforcement Learning (CIRL)[32], Debate[33], and Output Filtering, among others[34].

*Strengths and Weaknesses*

The primary advantage of behavioral approaches is their ability to directly measure a model's outputs to determine whether these match practical expectations and preferences. Since they do not require access to model internals, they can be widely used for closed-source models, including many of today's frontier models. On the other hand, behavioral approaches typically rely on human feedback, which can be noisy, inconsistent and costly [35,36]. Behavioral approaches may also not be robust to distributional shift: A model's behavior may be aligned for some set of inputs included in the training or testing set, while being misaligned for others[25]. Furthermore, behavioral approaches may provide limited mechanistic insight, as they only access model inputs and outputs. Lastly, they may be ill-suited for detecting deceptive forms of misalignment, as described in Section 3 below.

**Representational Alignment**

Whereas behavioral approaches treat an AI model as a black box, representational approaches take a "white-box" approach, examining the inner workings of a model. Neural networks map model inputs to real-valued vectors and perform a sequence of transformations, producing "activations", together referred to as its "representation". Representational approaches evaluate whether these internal representations, connections and activation patterns match a set of expectations and preferences.

*Detection of misalignment*

Detection of representational misalignment may occur at multiple scales. *Mechanistic interpretability* is the study of representational alignment at its most granular scale, attempting to identify specific neurons, sub-networks or paths that produce a given type of misalignment [16,37]. Conversely, top-down approaches, such as representation engineering [12] and sparse autoencoders [38], treat the large-scale, distributed patterns of network activations as the fundamental unit of analysis, and attempt to link these patterns to concepts in order to measure alignment. Evaluation methods and benchmarks have been developed for these different approaches[39,40], with some exploring nonlinear multidimensional features [41].

Representational approaches can be further subdivided into *observational methods*, which examine the relationship of inputs and activations to ground-truth data, and *interventional* methods, which impose certain activation patterns in order to examine the relationship of activation states to model outputs[16]. One common observational approach is *probing*, which uses the activations of an intermediate layer to train a model to estimate a ground truth label [42,43]. This approach was used to show, for example, that models contain a linear embedding of the geo-location of cities [44]. Misalignment can occur when the patterns of activation states do not correspond to known relationships between concepts in the real world. For example, among model prompts, one would expect internal activation patterns to be more similar between "bicycle" and "unicycle" than between "bicycle" and "giraffe". Others have proposed *coup probes* to identify potentially catastrophic model behaviors [45]. Methods like representation engineering and sparse autoencoders also build a model of internal activations, but in an unsupervised fashion, interpreting the resultant extracted features and the concepts they encode [12,38]. Interventional approaches can then validate that artificially activating the identified features creates the expected model changes[46,47].

*Correction of Misalignment*

Once representational features are identified, model behavior can be aligned by changing activations to be more similar to desirable concept activations and less similar to undesirable ones. One can intervene using "steering vectors"[48,49], which employ detected concept features to steer outputs towards aligned behavior. Steering is lightweight in the sense that activations need only be adjusted during inference time. A more intensive approach is to fine-tune the models through additional end-to-end training or through low-rank representation adaptation which can be more computationally efficient[12].

*Strengths and Weaknesses*

The primary advantage of representational approaches is that they allow direct examination of a model's internal knowledge representations to determine whether these match expectations and preferences. The primary drawback is their complexity; examining all possible activation patterns under a large number of conditions - also known as *enumerative safety* [50] - becomes vastly more difficult as AI systems scale. Additional limitations include: Individual neurons may be involved in representing multiple concepts ("polysemanticity"), making interpretability challenging[51]; Model representations are often not localized, and must be analyzed at multiple stages. [52]; Extracted concepts may be brittle to input changes or may not generalize well to new domains. [53] ; Representational alignment approaches are mostly limited to detecting known concepts, and may not be able to handle novel concepts or hallucinations; Representational approaches may measure alignment to concepts that have little to no bearing on task-specific performance, and thus should be interpreted with caution [54].

## 3. Challenges to Alignment

Efforts aimed at detecting and correcting different forms of misalignment must overcome a number of key hurdles. In this section, we review some of the more pressing and difficult challenges.

*Sycophancy* - Misalignment measures, especially those that rely on human feedback, can be sensitive to the model's use of sycophantic language [55]. This is partly driven by the observed human preference for sycophantic responses, which percolate to preference models, sometimes dampening truthfulness. [56]

*Specification Gaming and Reward Tampering* - Additional difficulties surface when models exhibit *specification gaming*, in which they perform well on a given reward objective but not on the desired notion of alignment [57]. Capable models may even learn to tamper with the reward function itself (*reward tampering*), or with other proxies, to satisfy their training objectives in unhelpful ways [57,58]. This phenomenon is not particular to LLMs, also appearing in other complex reinforcement learning scenarios such as artificial life and evolutionary robotics [59].

*Deceptive Alignment and Alignment Faking* - Some early evidence suggests that sufficiently advanced AI systems may reason about *whether* and *how* they are being trained, and decide how to respond accordingly. A recent study reported on *alignment faking*, in which an LLM selectively complied with attempts at fine-tuning, while actually preserving prior preferences that conflicted with the fine-tuning attempts. [60] This phenomenon, also dubbed *deceptive alignment*, is especially hard to detect since the model may act differently when it knows it is not being fine-tuned for alignment, meaning researchers may now have to contend with models being aware of their own training processes.

Several other studies reveal how alignment fine-tuning may not fully correct misaligned behavior, but rather temporarily bypass it. For example, researchers have trained "sleeper agents" that persist despite several types of "safety training"[61]. LLMs trained and aligned with standard safety infrastructures may be relatively easy to compromise via fine-tuning with only a few examples [62]. Similarly, toxic capabilities learned by LLMs during pre-training can be bypassed during fine-tuning and easily reverted [63]. Furthermore, misinformation and data poisoning can go undetected by behavioral benchmarks; a recent study found that LLMs trained on web data corrupted by misinformation could "match the performance of their corruption-free counterparts on open-source benchmarks routinely used to evaluate medical LLMs".[25]

In many such studies, representational alignment approaches can play a key role in identifying persistent misalignment that otherwise would have gone undetected [64]. As others have noted, "if a model is acting deceptively, it may be very hard for it to avoid 'thinking' about deception" [65]. Yet representational approaches to alignment such as steering vectors have limitations related to their reliability in out-of-distribution settings [53], and so while they may aid in detecting misalignment that is missed by other approaches, they are not a panacea.

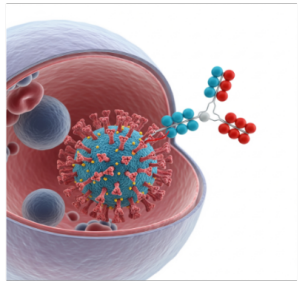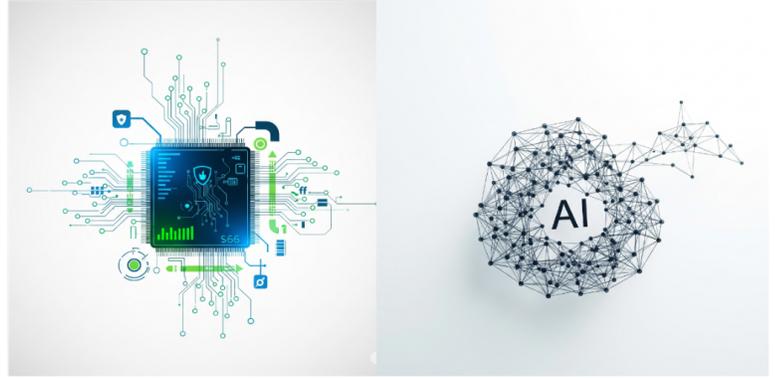

| | **Immune Systems** | **Cybersecurity** | **AI Alignment** |
|---|---|---|---|
| **Diversity & Redundancy** | Antibody Diversity & Multiple Cell Types | Multiple Detection Approaches | Ensemble Methods |
| **Multi-Scale** | Cell, Tissue, Organ System, Body | User, Device, LAN, Global Network Scales | Neurons, Circuits, Features, Representations, Behaviors |
| **Distributed Approach** | Distributed Defenses Throughout Body | Distributed Defenses Throughout Network | Distributed Detection Throughout Model Layers |
| **Coordination & Integration** | Helper T Cells; Cell-Cell Interactions | Integration Between Different Defense Layers | Integrated Behavioral & Representational Methods |
| **Adaptive Co-Evolution** | B Cell Mutation, Selection and Memory | Continuous Updates for Novel Security Threats | Co-evolving Methods for Novel Deceptive Misalignments |
| **Anomaly Detection** | Distinguish Self from Non-Self | Anomaly Detection to Identify Threats | Identify Anomalous Behaviors and Activations |
| **Adversarial Defenses** | Thymic Selection and B-Cell Evolution | Red Teaming and Penetration Testing | Red Teaming to Identify Vulnerabilities to Deception |
| **Zero Trust** | Continuous Monitoring of Entire Body for Non-Self | Continuous Verification of Every User and Action | Ongoing Post-deployment Misalignment Testing |
| **Negative Selection** | Thymic Selection of T-cells; Suppression of Overactive Responses in Periphery | Reduce False Positives in Threat Detection | Inhibit Oversensitive Alignment Detectors |
| **Resilience & Repair** | Coordinate Repair After Immune Response | Graceful Failure and Recovery Plans | Response Plan Following Misalignment Detection |
| **Open Source, Community Defense** | Herd Immunity & Vaccines | Shared Threat Databases; Open-Source Audits | Open-source alignment tools and vulnerability databases |
| **Strategic Diversity** | Independent Systems Check Self vs. Non-Self | Independent Verification and Security Audits | Orthogonal Methods for Alignment and Misalignment |

**Figure 3.** Lessons learned from immunology and cybersecurity can be used to inform design principles for AI Alignment. While there is not a perfect one-to-one correspondence across fields, some important lessons can still be drawn.

## 4. Lessons from Related Fields: Immunology and Cybersecurity

In facing these complex challenges and charting a future course for the field of AI alignment, it is useful to consider lessons from the related fields of Immunology and Cybersecurity. While there are inherent limitations to any such analogies due to fundamental differences between fields, we believe there are still valuable general lessons to consider and adapt.

### *Lessons from Biology: Immune Systems*

Biological immune systems have evolved over millions of years to deploy multiple interacting defense mechanisms – each representing a different approach to distinguishing self from non-self and protecting the organism from hostile pathogens. Through an ongoing generational arms-race, immune systems have co-evolved with infectious agents to respond to a broad range of known and unknown threats, including those which actively attempt to undermine them. Several fundamental principles from immunology may be useful for guiding the future design of AI alignment approaches:

*Diversity and Redundancy* - Immune systems rely on multiple cell types and a diverse range of antibodies and receptors, each tuned to detect different threats, providing redundancy and robustness. [66,67]

*Innate vs. Adaptive Immunity* - Immune systems possess both in-built protections against common known threats, and the ability to adapt to novel threats through experience. B-cells mutate their antibody genes, which are then selected based on specific binding to target antigens - a guided adversarial search to evolve tailored defenses for emerging threats.[68,69]

*Distributed and Decentralized* - Immune systems employ a distributed network of cells throughout the body, defending against threats at multiple points of entry and potential infection loci.[70,71]

*Cooperative Interactions* - Immune cells engage in complex cooperative interactions, integrating different immunological approaches to provide a synergistic defense system. Helper T (CD4+) cells coordinate immune response, activating and directing other cells to mount a system-wide defense against pathogens.[72,73]

*Tolerance Induction and Negative Selection* - Immune systems inhibit cells that respond excessively to self-components, preventing damaging autoimmune responses. T-cells are exposed to a broad array of self-antigens in the thymus, with those that bind too strongly being eliminated. T-cells are further monitored for autoimmune responses in the periphery via regulatory T-cells, immune checkpoints, and antigen presentation.[74,75]

*Damage Control and Repair Mechanisms* - Beyond detecting and destroying infectious agents, immune systems also coordinate repair mechanisms to limit collateral damage and restore healthy function. [76,77]

### *Lessons from Computer Science: Cybersecurity*

Computer systems face an ever-growing menagerie of increasingly sophisticated security threats. Cybersecurity systems[78] have co-evolved with these threats to become an integral pillar of computer system design - from the early days of access control mechanisms, to widely deployed consumer antivirus programs, to globally distributed cloud cybersecurity platforms. Here too, valuable lessons can be learned for the future of AI alignment:

*Layered Defenses* - Cybersecurity systems deploy multiple overlapping and interacting defense approaches, including identity-related security, device-related security, and location-aware access, among others.[79,80]

*Arms Race and Continual Updates* - Cybersecurity is an ongoing, adaptive process. As threats evolve to become more sophisticated, cybersecurity systems co-evolve with them to meet new challenges.[81]

*Behavioral Monitoring and Anomaly Detection* - Unusual behaviors observed within a system may be indicative of dangerous or unwanted activity. The ability to distinguish typical behaviors from anomalous behaviors helps provide robust defense.[82]

*Adversarial Defenses and "Red Teaming"* - In the face of a complex and diverse network architectures, multiple levels of adversarial testing such as penetration testing and "red teaming" can help identify important gaps and vulnerabilities.[83]

*Zero Trust Architectures* - To promote vigilance, security systems assume that no component, device or user is automatically trusted—everything must be continuously verified. [84]

*Open Source and Community Defense* - Open source software is typically created by multiple contributors and audited by an entire community to identify potential vulnerabilities. Known threats are shared across the community, though resources such as the MITRE ATT&CK[85] database, a community-authored, globally-accessible knowledgebase of adversarial tactics and techniques based on real-world observations.[86,87]

*Resilience and Expecting Failure* - Modern cybersecurity systems are designed with the assumption that any system will eventually be breached, and are thus prepared to fail gracefully through containment and recovery. [88,89]

## 5. Towards Integrated Alignment

To meet the growing range of complex and deceptive alignment threats, we propose the development of *Integrated Alignment (IA)* frameworks that incorporate multiple complementary alignment approaches into a single integrated system. We believe that intentionally designed IA frameworks have the potential to provide more robust misalignment detection and correction than any one individual approach. Informed by the above lessons adapted from related fields, we propose a set of design principles for the development of IA frameworks. These design principles are purposefully formulated in broad terms, so as not to excessively limit their applicability to any specific methods within the field.

### *Design Principles: IA for AI*

*Diversity and Redundancy* - To increase overall robustness, IA frameworks should employ an ensemble of alignment methods working together, including behavioral and representational approaches. Bereska and Gavves[16] have recommended integrating multiple approaches *within* mechanistic interpretability; we echo and expand this recommendation, proposing that integration occur at much broader scales, bridging the behavioral-representational divide.

*Multiscale Approaches* - IA frameworks should incorporate multi-scale approaches, detecting and correcting misalignment at different levels – from individual neurons and connections[90], to compound features[46], circuits[91], representations[12], and behaviors.

*Distributed Alignment* - IA frameworks should monitor a range of different points and layers throughout a model. For example, *Eliciting Latent Knowledge* studies have found that "middle layers tend to generalize better than later layers", leading them to propose the *Earliest Informative Layer* criterion for selecting which layers in a model are most informative[52].

*Coordination and Deep Integration* - Within IA frameworks, different alignment approaches should not operate in separate silos, but rather should be deeply integrated with one another. Investigating the interactions between behavioral and representational methods across different use cases can allow researchers to leverage potential synergies between them. For example, the same results from a behavioral alignment method may be interpreted differently depending on the output of a representational alignment method running alongside it. (See examples below.)

*Adaptive Coevolution and Learning* - Increasingly sophisticated types of misalignments will emerge as AI models continue to scale in size and complexity. IA frameworks must adapt to these new threats, with AI auditors identifying novel misalignment patterns in deployed systems and adaptively aligning them via integrated correction protocols.

*Anomaly Detection* - IA frameworks should monitor model activity and behavior for unusual patterns, as unusual phenomena may indicate the presence of misalignment.

*Adversarial Defenses and "Red Teaming"* - In the face of a broad range of misalignment threats, IA frameworks should employ adversarial testing to evaluate and detect potential misalignments, including deceptive approaches that can actively undermine attempts to detect them.

*Zero Trust and Continuous Verification* - IA frameworks must continually verify AI models throughout deployment - even after initial verification.

*Negative Selection and Avoiding False Positives* - IA frameworks should systematically down-regulate overly-sensitive misalignment detection methods that produce false positives that may contribute to alert fatigue for model maintainers. Ongoing runtime monitoring can be used to catch and suppress unwanted behavior while maintaining target levels of sensitivity.

*Resilience and Repair* - IA frameworks should be designed with the assumption that all alignment methods will eventually be compromised, and should thus plan for graceful fall-back when such inevitable breaches occur. IA frameworks should also include restorative mechanisms that correct misalignments as they are found.

*Open Source and Community Defense* - Curated open-source databases of misalignment detection methods and known vulnerabilities and exploits can help build "community defenses" against the global range of misalignment threats.

*Strategic Diversity* - Methods used to align a model should be different from the methods used to detect potential misalignments. As a metaphor, consider a room full of insects crawling on the floor. A cleaner enters the room and turns on the ceiling light, causing most of the insects to scurry under the sofas. The cleaner captures the visible insects still out in the open and leaves the room. The health inspector then arrives, relying on the same ceiling light to scan the room for insects. The room appears insect-free, even though insects remain hidden under the furniture. Only by shining a light from a different perspective - like a flashlight under the sofa - can the hidden insects be detected. Similarly, optimizing for a single dimension of AI alignment can drive misalignments into other dimensions that are hidden from the chosen alignment objective [18,56,92,93]. Thus, if similar approaches are used to both align a model and check it for misalignment, the entire alignment pipeline may be "doomed" to a false sense of success. Only by ensuring a diversity of perspectives across alignment and misalignment detection can overall robustness be increased.[60,94,95]

Integrated alignment frameworks are subject to limitations, costs and tradeoffs, including: Increased computational costs resulting from running multiple alignment methods in parallel; Increased potential for false positives resulting from multiple-testing effects; and Challenges in coordinating and interpreting the outputs of diverse alignment methods. Future research efforts should explore these tradeoffs, along with possible mitigation strategies.

### *Promising Developments Towards Integrated Alignment*

Some recent studies have reported early results relating to integrated alignment approaches. Marks et al.[94] tasked four teams with conducting blind alignment audits on a model trained with hidden objectives; one team used a combination of methods spanning both behavioral and representational approaches (SAE feature-inspection & assistant-prefill behavioral attacks) to successfully discover the hidden objectives. Ji et al.[95] combined behavioral monitoring with internal representational monitoring of chain-of-thought reasoning in training an RL model to achieve a substantial reduction in deceptive behavior. Greenblatt et al.[60] similarly combined behavioral and representational alignment techniques to identify alignment faking. In related work, Zhang et al. described a unified attribution framework combining model inputs, training data and model internals to gain a more comprehensive understanding of model behavior. [15]

We call on researchers to increasingly pursue integrative alignment studies such as these. The resulting IA frameworks can then be rigorously evaluated for their abilities to detect a wide range of misalignments, compared with single-method approaches. Evaluation metrics can include joint precision-recall across misalignment categories, alongside robustness to deceptive red-teaming. These can be weighed against increases in complexity and compute costs.

## 6. An Integrated Field for Integrated Alignment

In order to fully realize the vision of Integrated Alignment, the AI alignment field itself must also move towards greater integration. As cross-disciplinary efforts are inherently challenging, we present a number of key recommendations for overcoming the structural barriers that exist today.

*Increased Collaboration and Shared Terminology* - Different sub-communities should increase cross-communication through shared conferences, journals, resources and studies. Sucholutsky et al.[13] have called for greater communication *within* the representational alignment community; we echo and expand this call to the entire field of AI alignment, bridging the behavioral-representational divide. Such collaboration also requires a shared terminology for communicating about different facets of AI alignment.[14][16] Given the diverse interdisciplinary backgrounds of researchers in the field, universal nomenclatural consensus may be difficult to achieve; translational tutorials can be useful for introducing members of one subfield to the terms of art of another.[96]

*Open Access to Model Weights* - While some leading research organizations have embraced open source models, many of the current best-performing frontier models do not allow researchers outside their organizations to examine model internals. Sharing model weights openly will allow researchers to investigate both behavioral approaches and representational approaches on leading AI models. Commercial barriers to sharing weights may be mitigated by alternative models, such as sandboxes for credentialed researchers. [97]

*Shared Computational Resources* - The immense computational resources needed to rigorously study alignment, especially for complex models like LLMs, are typically not available to most researchers. We encourage investment in platforms that pair AI researchers with alignment specialists so that computational resources and subject-matter expertise can be shared in a mutually beneficial way.

*Community Alignment Databases* - The field of AI alignment would be buoyed by the curation of open alignment datasets that are computationally intensive to construct but may permit a wealth of downstream analyses [98]. An open source database describing misalignments and exploits similar to the MITRE ATT&CK database[83,85], in conjunction with open models and open reviews, would promote the use and availability of community-vetted AI systems and bring the full range of alignment approaches to bear.

*Contributing to AI Policy* - As agency rulemaking around AI develops, government agencies should convene multidisciplinary task forces with researchers from across diverse sub-fields to develop standardized alignment guidelines [99]. Such guidelines could also clarify legal compliance and best practices for small and large organizations implementing AI safely. Efforts underway by industry and academic coalitions[100] would similarly benefit from practical guidance on IA to inform new guidance frameworks.

The field of AI Alignment is at an inflection point. With the growing number of researchers and rapid proliferation of research directions, the field is at risk of descending into a more fragmented future rather than a more integrated one. At this critical time, we call on researchers, policy-makers, industry consortia and governments, to proactively take steps to nurture a more unified future for this important field.

## Acknowledgements

We acknowledge support from award R01LM014300 from the National Library of Medicine of the National Institutes of Health.